\documentclass{iau} 
\usepackage{graphicx}
\usepackage{natbib}
\usepackage{hyperref}

\title[Gravitational few-body problems] 
{Modeling gravitational few-body problems with \textsc{TSUNAMI} and \textsc{OKINAMI}}

\author[Alessandro A. Trani]   
{Alessandro A. Trani$^{1,2}$ \& Mario Spera$^{3,4,5}$}

\affiliation{$^1$Department of Earth Science and Astronomy, College of Arts and Sciences, The University of Tokyo, 3-8-1 Komaba, Meguro-ku, Tokyo 153-8902, Japan \\
$^2$Okinawa Institute of Science and Technology, 1919-1 Tancha, Onna-son, Okinawa 904-0495, Japan \\ $^3$ SISSA, Via Bonomea 265, I-34136, Trieste, Italy \\
$^4$ INFN, Sezione di Trieste, I-34127 Trieste, Italy \\
$^5$ INFN, Sezione di Padova, Via Marzolo 8, I–35131, Padova, Italy \\ email: {\tt aatrani@gmail.com}}

\pubyear{2015}
\volume{xxx}  
\jname{Title of your IAU Symposium}
\editors{A.C. Editor, B.D. Editor \& C.E. Editor, eds.}

\begin{document}

\maketitle

\begin{abstract}
In recent years, an increasing amount of attention is being paid to the gravitational few-body problem and its applications to astrophysical scenarios. Among the main reasons for this renewed interest there is large number of newly discovered exoplanets and the detection of gravitational waves. 
Here, we present two numerical codes to model three- and few-body systems, called \textsc{tsunami} and \textsc{okinami}. The \textsc{tsunami} code is a direct few-body code with algorithmic regularization, tidal forces and post-Newtonian corrections. \textsc{okinami} is a secular, double-averaged code for stable hierarchical triples. 
We describe the main methods implemented in our codes, and review our recent results and applications to gravitational-wave astronomy, planetary science and statistical escape theories. 

\end{abstract}

\firstsection 
\section{Introduction}

The gravitational three-body problem has a 300 years old history, dating back to Newton, Poincar\'e and many others. Rather than just being a didactic tool for the mathematical physicist, the three-body problem has numerous applications to modern astrophysical conundrums. Thanks to the recent advancement in observational astronomy, the three-body problem (and more generally, the few-body problem) is experiencing a renewed interest. 
Driving such interest is the detection of gravitational waves in 2015, and the subsequent birth of gravitational-wave astronomy \citep{gwtc-3}. In fact, three-body interactions between compact objects have been proposed as one of the key formation mechanisms of gravitational-wave sources.

Another area of interest for three-body problems is exoplanet formation and evolution. This was made possible thanks to the rapid increase in exoplanet detections from transit surveys \citep[K2, TESS,][]{k2mission,tess}, and the characterization of numerous exotic planetary systems (i.e. hot Jupiters, ultra-short period planets, compact resonant chains). The formation of such exotic systems can be explained with gravitational few-body interactions between planets or passing stars. In addition, the recent reports of exomoon candidates have opened up questions on how extrasolar moons form and evolve.

Modeling few-body gravitational interactions is not an easy task. One issue arises from the nature of the gravitational force, which scales as $\propto r^{-2}$, where $r$ is the separation between two particles. When two particles get very close, $r\rightarrow 0$ and the acceleration increases dramatically. Using traditional integrators, like the Runge-Kutta or Hermite methods, as the acceleration increases, the timestep needs to be reduced accordingly, in order to time-resolve the trajectory of the particles with sufficient accuracy. This can possibly lead to the halt of the
integration, or to the faster accumulation of integration errors due to the increased number of timesteps. 

We have developed two codes, named \textsc{tsunami} and \textsc{okinami} that employ different techniques in order to accurately model few-body gravitational interactions. Here, we describe the main numerical methods that we implemented, along with their applications to astrophysical scenarios.

\section{Overview of the codes}
\textsc{tsunami} and \textsc{okinami} implement different methods and therefore have slightly different scopes. The main difference is that while \textsc{tsunami} can simulate systems of hundreds of particles in arbitrary configurations, \textsc{okinami} can model only hierarchical stable triples. Both codes can be interfaced through a dedicated Python library and come with several example scripts. 

\subsection{The \textsc{tsunami} code}
\textsc{tsunami} is based on the following techniques: regularization of the equations of motion, chain coordinates to
reduce round-off errors and Bulirsch--Stoer extrapolation. The first technique (regularization) takes care of the singularity of the gravitational potential for $r\rightarrow 0$. The second technique (chain coordinates) helps reducing the round-off errors in hierarchical systems, which arise with the center-of-mass coordinates, without the need to include numerically expensive techniques of compensated summation. The third method (Bulirsch--Stoer extrapolation) increases the accuracy of the integration and makes it adaptable over a wide dynamical range.

\textsc{tsunami} solves the Newtonian equations of motion derived from a modified, extended Hamiltonian \citep{mik99a,mik99b}. As a consequence, time is another variable that is integrated along positions and velocities of the particles. Along one timestep of fictitious time $\Delta S$, the physical time $\Delta T$ is advanced by:
\begin{equation}\label{eq:ds}
	\Delta T = \frac{\Delta S}{\alpha U + \beta \Omega + \gamma}
\end{equation}
where $U$ is the potential energy, $\Omega$ is a function of positions, and $\alpha$, $\beta$ and $\gamma$ are arbitrary coefficients. Setting the values for $(\alpha, \beta, \gamma)$ effectively changes the regularization algorithm. $(\alpha, \beta, \gamma) = (1,0,0)$ corresponds to the logarithmic Hamiltonian algorithm, $(\alpha, \beta, \gamma) = (0,1,0)$ is equivalent to the time-transformed leapfrog scheme, and for $(\alpha, \beta, \gamma) = (0,0,1)$ the integration scheme reduces to the non-regularized leapfrog.

Positions and velocities are integrated in a chain coordinate system, rather than in the center-of-mass coordinates. This has the effect of reducing by 1 the number of equations to be integrated, and more importantly it reduces round-off errors when calculating distances between close particles far from the center of mass of the system \citep{mik93}.
These errors can quickly arise if the inter-particle separation
is very small compared to the distance from the center of mass, due to the limits of floating-point arithmetic, which can happen, for example, in case of close binaries far from a massive black hole.
The chain of inter-particle vectors is formed so that all particles are included in the chain. The first segment of the chain is chosen to be the shortest inter-particle distance in the system. The next segment is included so that it connects the particle closest to one of the ends of the current chain. This process is repeated until all particles are included.
As the system evolves, care is taken to update the chain so that any chained vector is always shorter than adjacent non-chained vectors. It is possible to directly transform the old coordinates into new chain coordinates without passing through the center-of-mass coordinates.

Finally, a simple leapfrog integration might not be accurate enough for some applications. Therefore, the accuracy of the integration can be improved with the Bulirsch-Stoer extrapolation.
The idea behind Bulirsch–Stoer extrapolation is to consider the results of a numerical integration as being an analytic function of the stepsize $h$. The solution of a given time interval $\Delta S$ is computed for smaller and smaller substeps $h = \Delta S / N_{\rm steps}$ and then it is extrapolated to $h\rightarrow 0$, using rational or polynomial functions.

The above integration scheme works well for Newtonian gravity. However, this is not enough to model some systems like binary black holes or planets, which require additional physics. \textsc{tsunami} implements additional forces, like equilibrium tides \citep{hut81}, dynamical tides \citep{sam18a} and post-Newtonians corrections of order 1, 2 and 2.5, using the midpoint step described in \citet{mik08}.

\begin{figure}
	\begin{center}
		\includegraphics[width=\textwidth]{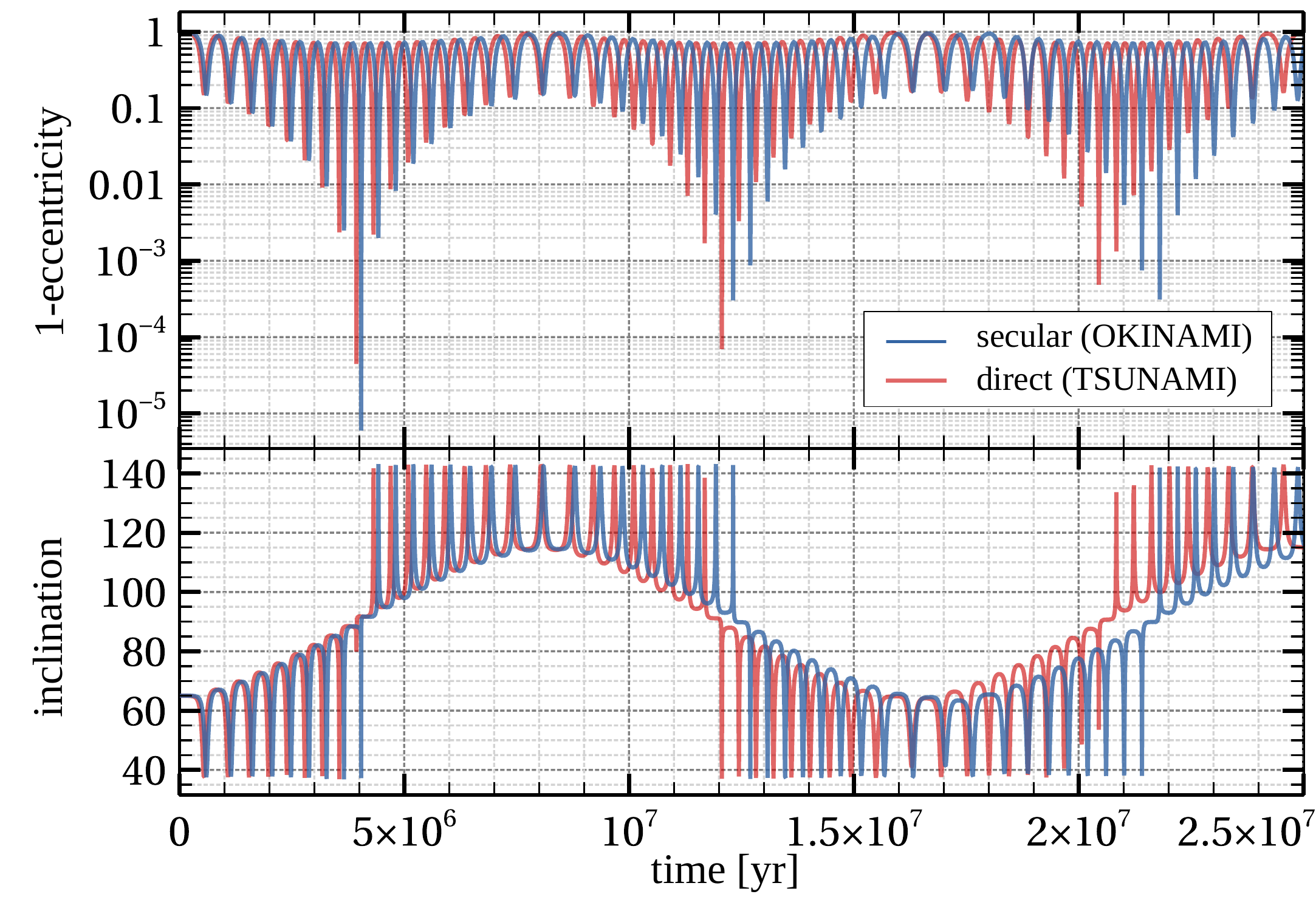} 
		\caption{Comparison between \textsc{tsunami} (red, direct integration) and \textsc{okinami} (blue, secular average) in evolving a Jupiter-Sun-brown dwarf system. The top panel shows the eccentricity of the Jupiter's orbit around the Sun-like star, while the bottom panel shows the mutual inclination between the Jupiter's orbit and the brown dwarf's orbit. As the Jupiter is perturbed by the outer brown dwarf, the Jupiter exhibits von~Zeipel-Kozai-Lidov oscillations, with the typical flip of the orbit associated to the octupole-level secular interaction. This figure reproduces fig.3 from \citep{naoz2013a}.}
		\label{fig1}
	\end{center}
\end{figure}

\subsection{The \textsc{okinami} code}
Unlike \textsc{tsunami}, \textsc{okinami} is limited to stable hierarchical triples, that is, a binary whose center of mass forms another binary with a tertiary body. 
At its core, \textsc{okinami} integrates the equations of motion derived from a three-body Hamiltonian, expanded at the octupole-level interaction and averaged over the mean anomalies of the inner and outer orbits. The double-average has the advantage of considerably speeding up the integration, because it avoids the integration of the ``fast angles''. On the other hand, this has two consequences: the information about the individual positions of the bodies along their orbit is lost, and the equations cannot describe the evolution of the system on timescales shorter than the inner and outer orbital periods.
After the double average, what we obtain is a set of ordinary differential equations for the inner and outer eccentricities, ($e_1$, $e_2$), arguments of pericenter ($\omega_1$, $\omega_2$), longitudes of the ascending nodes ($\Omega_1$, $\Omega_2$) and orbital inclinations ($i_1,i_2$). \textsc{okinami} integrates these equations using and adaptive Runge-Kutta-Fehlberg of order 7. In addition, \textsc{okinami} implements also equilibrium tides and post-Newtonian terms of orders 1, 2 and 2.5 for the inner orbit.

As an example, \autoref{fig1} shows the evolution of inclination and eccentricity of the orbit of a Jupiter-sized planet around a Sun-like star, orbited by a distant brown dwarf. The integration with \textsc{tsunami} takes about 5 minutes, while the one with \textsc{okinami} takes less than a second.

\section{Applications}

\subsection{Exoplanets and exomoons}
In \citet{2020MNRAS.499.4195T} we investigated the fate of exomoons around migrating hot Jupiters. We considered the scenario in which hot Jupiters experience high-eccentricity, tidally driven migration, due to the gravitational perturbation from a distant companion star. Physically, the system is a 4-body problem composed of two nested hierarchical triple systems: one triple is composed of the primary star, companion star and the Jupiter, the second triple is constituted by the moon, its host Jupiter and the main star.

This kind of system is an ideal test bench for \textsc{tsunami}. Our code can accurately model the tidal forces required for the migration of the Jupiter and general relativity precession that can alter the long-term dynamics of the Jupiter and its moon.
We found that exomoons are unlikely to survive the migration process of the host Jupiter. Massive moons can prevent the migration process entirely, by
suppressing the eccentricity excitation induced by the secondary star.
If the moon cannot shield the planet from perturbations, the Jupiter's orbit becomes increasingly eccentric, triggering the dynamical instability of the moon. Subsequently, most exomoons end up being ejected from the system or colliding with the primary star and the host planet. Only a few escaped exomoons can become stable planets after the Jupiter has migrated, or by tidally migrating themselves. 

Even though close-in giants are ideal candidates for exomoon detections, our results suggest that it is unlikely for exomoons to be discovered around them, at least for planets migrated via high-eccentricity tidal circularization. Nonetheless, tidally disruptions or collisions of exomoons can still leave observational signatures, such as debris disks or chemically altered stellar atmospheres.

Besides exotic scenarios like exomoons and hot Jupiters, \textsc{tsunami} is an excellent tool to assess the stability and long-term evolution of planetary systems \citep[e.g.][]{2019MNRAS.484....8L}.

\subsection{Gravitational-wave radiation sources}

The astrophysical origin of gravitational-wave events from coalescing black holes binaries is still debated, though many of the proposed formation scenarios involve some kind of few-body gravitational interaction.
Specifically, chaotic three-body interactions between compact-object binaries and single black holes can happen frequently in dense stellar systems. These interactions can alter the spin-orbit orientation of black hole binaries, which can then be inferred from gravitational-wave observations. 

In \citet{2021MNRAS.504..910T}, we estimated the spin parameter distributions of merging black-hole binaries, comparing them with the currently available data. Here, we introduced a new formation scenario that combines elements from both the isolated and the dynamical formation scenarios. We ran an extensive set of highly-accurate simulations with \textsc{tsunami}, and we used the results to estimate the intrinsic merger rates of black-hole binaries in combination with a semi-analytic model.

Assuming low natal black-hole spins ($\chi < 0.2$), our scenario reproduces the distributions of $\chi_{\rm eff}$ and $\chi_{\rm p}$ inferred from current observations. In particular, this model can explain the peak at positive $\chi_{\rm eff}$ with a tail at negative $\chi_{\rm eff}$, and the broad peak at $\chi_{\rm p} \sim 0.2$.
This is in sharp contrast with the predictions of the isolated and the dynamical scenarios: the first fails to produce negative $\chi_{\rm eff}$, while the second predicts a symmetric distribution around $\chi_{\rm eff} \sim 0$. 

In \citet{2021arXiv211106388T} we examined merging compact-object binaries in hierarchical triple systems. We first obtained a large sample of triples formed in low mass clusters through dynamical interactions, simulated using direct N-body methods \citep{rastello2021}. Because we selected only stable triples, we evolved them using \textsc{okinami}.
We obtained the merger properties of binary black holes, black hole--neutron stars, and black hole--white dwarfs. The rates for binary black holes, black hole--neutron stars are about 100 times lower than those of binary mergers
from the same clusters. This is caused by the lower merger efficiency of triple systems, which is about 100 times
lower than that of binaries.
Nonetheless, compact objects merging from triples have unique properties that can be used to discriminate them from other formation channels.

Compared to binary black-hole mergers from open clusters, mergers from triples have more massive primaries, with a mass distribution peaking at around $30 \rm\,M_\odot$ rather than $10 \rm\,M_\odot$. The mass ratio also peaks at smaller values of $0.3$, in contrast to the cluster binaries pathway, which favors equal-mass binaries. This is caused by the
von~Zeipel-Kozai-Lidov mechanism, whose eccentricity-pumping effect is enhanced at
low mass ratios.

\textsc{tsunami} is also ideal to model compact objects and stars in proximity to massive black black holes. We investigated the impact of three-body encounters around massive black holes on binary black hole coalescence in \citet[see also \citealt{2019ApJ...875...42T} and \citealt{2020IAUS..351..174T}]{2019ApJ...885..135T}

\subsection{Statistical solutions to the three-body problem}
The gravitational three-body problem is chaotic  has no general analytic solution, and only partial statistical solutions have been achieved so far \citep[see][and references therein]{stone2019}. 
The main idea behind these statistical escape theories is to leverage chaos to predict the evolution of a three-body problem only in a statistical sense, using the assumption of thermodynamical ergodicity. 
Recently, \citet{kol2021} introduced a novel statistical theory based on the flux of phase space, rather than on phase-space volume like all the previous theories. 

In a series of papers, we have been testing the statistical theories by simulating a large ensembles of three-body systems with \textsc{tsunami} and comparing the final outcome distributions to the theoretical predictions.
Our results in \citet{2020MNRAS.497.3694M} and \citet{2021MNRAS.506..692M} show that the flux-based theory is in tighter agreement with the outcome of three-body simulations, compared to the previous statistical theories. 
We are now in the process of further testing the potential of this theory with the aim of providing a complete, accurate statistical description of the three-body problem.

\bibliographystyle{astron}

\bibliography{totalms,addms}

\begin{thebibliography}{}

\bibitem[\protect\astroncite{{Howell} et~al.}{2014}]{k2mission}
{Howell}, S.~B., {Sobeck}, C., {Haas}, M., {Still}, M., {Barclay}, T.,
  {Mullally}, F., {Troeltzsch}, J., {Aigrain}, S., {Bryson}, S.~T., {Caldwell},
  D., {Chaplin}, W.~J., {Cochran}, W.~D., {Huber}, D., {Marcy}, G.~W.,
  {Miglio}, A., {Najita}, J.~R., {Smith}, M., {Twicken}, J.~D., and {Fortney},
  J.~J.: 2014,
\newblock {\em \pasp} {\bf 126(938)}, 398

\bibitem[\protect\astroncite{{Hut}}{1981}]{hut81}
{Hut}, P.: 1981,
\newblock {\em \aap} {\bf 99}, 126

\bibitem[\protect\astroncite{{Kol}}{2021}]{kol2021}
{Kol}, B.: 2021,
\newblock {\em Celestial Mechanics and Dynamical Astronomy} {\bf 133(4)}, 17

\bibitem[\protect\astroncite{{Livingston} et~al.}{2019}]{2019MNRAS.484....8L}
{Livingston}, J.~H., {Dai}, F., {Hirano}, T., {Gandolfi}, D., {Trani}, A.~A.,
  {Nowak}, G., {Cochran}, W.~D., {Endl}, M., {Albrecht}, S., {Barragan}, O.,
  {Cabrera}, J., {Csizmadia}, S., {de Leon}, J.~P., {Deeg}, H.,
  {Eigm{\"u}ller}, P., {Erikson}, A., {Fridlund}, M., {Fukui}, A., {Grziwa},
  S., {Guenther}, E.~W., {Hatzes}, A.~P., {Korth}, J., {Kuzuhara}, M.,
  {Monta{\~n}es}, P., {Narita}, N., {Nespral}, D., {Palle}, E., {P{\"a}tzold},
  M., {Persson}, C.~M., {Prieto-Arranz}, J., {Rauer}, H., {Tamura}, M., {Van
  Eylen}, V., and {Winn}, J.~N.: 2019,
\newblock {\em \mnras} {\bf 484(1)}, 8

\bibitem[\protect\astroncite{{Manwadkar} et~al.}{2021}]{2021MNRAS.506..692M}
{Manwadkar}, V., {Kol}, B., {Trani}, A.~A., and {Leigh}, N. W.~C.: 2021,
\newblock {\em \mnras} {\bf 506(1)}, 692

\bibitem[\protect\astroncite{{Manwadkar} et~al.}{2020}]{2020MNRAS.497.3694M}
{Manwadkar}, V., {Trani}, A.~A., and {Leigh}, N. W.~C.: 2020,
\newblock {\em \mnras} {\bf 497(3)}, 3694

\bibitem[\protect\astroncite{{Mikkola} and {Aarseth}}{1993}]{mik93}
{Mikkola}, S. and {Aarseth}, S.~J.: 1993,
\newblock {\em Celestial Mechanics and Dynamical Astronomy} {\bf 57}, 439

\bibitem[\protect\astroncite{{Mikkola} and {Merritt}}{2008}]{mik08}
{Mikkola}, S. and {Merritt}, D.: 2008,
\newblock {\em \aj} {\bf 135}, 2398

\bibitem[\protect\astroncite{{Mikkola} and {Tanikawa}}{1999a}]{mik99a}
{Mikkola}, S. and {Tanikawa}, K.: 1999a,
\newblock {\em \mnras} {\bf 310}, 745

\bibitem[\protect\astroncite{{Mikkola} and {Tanikawa}}{1999b}]{mik99b}
{Mikkola}, S. and {Tanikawa}, K.: 1999b,
\newblock {\em Celestial Mechanics and Dynamical Astronomy} {\bf 74}, 287

\bibitem[\protect\astroncite{{Naoz} et~al.}{2013}]{naoz2013a}
{Naoz}, S., {Farr}, W.~M., {Lithwick}, Y., {Rasio}, F.~A., and {Teyssandier},
  J.: 2013,
\newblock {\em \mnras} {\bf 431(3)}, 2155

\bibitem[\protect\astroncite{{Rastello} et~al.}{2021}]{rastello2021}
{Rastello}, S., {Mapelli}, M., {di Carlo}, U.~N., {Iorio}, G., {Ballone}, A.,
  {Giacobbo}, N., {Santoliquido}, F., and {Torniamenti}, S.: 2021,
\newblock {\em arXiv e-prints} p. arXiv:2105.01669

\bibitem[\protect\astroncite{{Ricker} et~al.}{2015}]{tess}
{Ricker}, G.~R., {Winn}, J.~N., {Vanderspek}, R., {Latham}, D.~W., {Bakos},
  G.~{\'A}., {Bean}, J.~L., {Berta-Thompson}, Z.~K., {Brown}, T.~M.,
  {Buchhave}, L., {Butler}, N.~R., {Butler}, R.~P., {Chaplin}, W.~J.,
  {Charbonneau}, D., {Christensen-Dalsgaard}, J., {Clampin}, M., {Deming}, D.,
  {Doty}, J., {De Lee}, N., {Dressing}, C., {Dunham}, E.~W., {Endl}, M.,
  {Fressin}, F., {Ge}, J., {Henning}, T., {Holman}, M.~J., {Howard}, A.~W.,
  {Ida}, S., {Jenkins}, J.~M., {Jernigan}, G., {Johnson}, J.~A., {Kaltenegger},
  L., {Kawai}, N., {Kjeldsen}, H., {Laughlin}, G., {Levine}, A.~M., {Lin}, D.,
  {Lissauer}, J.~J., {MacQueen}, P., {Marcy}, G., {McCullough}, P.~R.,
  {Morton}, T.~D., {Narita}, N., {Paegert}, M., {Palle}, E., {Pepe}, F.,
  {Pepper}, J., {Quirrenbach}, A., {Rinehart}, S.~A., {Sasselov}, D., {Sato},
  B., {Seager}, S., {Sozzetti}, A., {Stassun}, K.~G., {Sullivan}, P.,
  {Szentgyorgyi}, A., {Torres}, G., {Udry}, S., and {Villasenor}, J.: 2015,
\newblock {\em Journal of Astronomical Telescopes, Instruments, and Systems}
  {\bf 1}, 014003

\bibitem[\protect\astroncite{{Samsing} et~al.}{2018}]{sam18a}
{Samsing}, J., {Leigh}, N.~W.~C., and {Trani}, A.~A.: 2018,
\newblock {\em ArXiv e-prints}

\bibitem[\protect\astroncite{Stone and Leigh}{2019}]{stone2019}
Stone, N.~C. and Leigh, N. W.~C.: 2019,
\newblock {\em Nature} {\bf 576(7787)}, 406

\bibitem[\protect\astroncite{{The LIGO Scientific Collaboration}
  et~al.}{2021}]{gwtc-3}
{The LIGO Scientific Collaboration}, {the Virgo Collaboration}, {the KAGRA
  Collaboration}, {Abbott}, R., {Abbott}, T.~D., {Acernese}, F., {Ackley}, K.,
  {Adams}, C., {Adhikari}, N., {Adhikari}, R.~X., {Adya}, V.~B., {Affeldt}, C.,
  {Agarwal}, D., {Agathos}, M., {Agatsuma}, K., {Aggarwal}, N., {Aguiar},
  O.~D., {Aiello}, L., {Ain}, A., {Ajith}, P., {Akcay}, S., {Akutsu}, T.,
  {Albanesi}, S., and {Allocca}, A.~a.: 2021,
\newblock {\em arXiv e-prints} p. arXiv:2111.03606

\bibitem[\protect\astroncite{{Trani}}{2020}]{2020IAUS..351..174T}
{Trani}, A.~A.: 2020,
\newblock in A. {Bragaglia}, M. {Davies}, A. {Sills}, and E. {Vesperini}
  (eds.), {\em Star Clusters: From the Milky Way to the Early Universe}, Vol.
  351, pp 174--177

\bibitem[\protect\astroncite{{Trani} et~al.}{2019a}]{2019ApJ...875...42T}
{Trani}, A.~A., {Fujii}, M.~S., and {Spera}, M.: 2019a,
\newblock {\em \apj} {\bf 875(1)}, 42

\bibitem[\protect\astroncite{{Trani} et~al.}{2020}]{2020MNRAS.499.4195T}
{Trani}, A.~A., {Hamers}, A.~S., {Geller}, A., and {Spera}, M.: 2020,
\newblock {\em \mnras} {\bf 499(3)}, 4195

\bibitem[\protect\astroncite{{Trani} et~al.}{2021a}]{2021arXiv211106388T}
{Trani}, A.~A., {Rastello}, S., {Di Carlo}, U.~N., {Santoliquido}, F.,
  {Tanikawa}, A., and {Mapelli}, M.: 2021a,
\newblock {\em arXiv e-prints} p. arXiv:2111.06388

\bibitem[\protect\astroncite{{Trani} et~al.}{2019b}]{2019ApJ...885..135T}
{Trani}, A.~A., {Spera}, M., {Leigh}, N. W.~C., and {Fujii}, M.~S.: 2019b,
\newblock {\em \apj} {\bf 885(2)}, 135

\bibitem[\protect\astroncite{{Trani} et~al.}{2021b}]{2021MNRAS.504..910T}
{Trani}, A.~A., {Tanikawa}, A., {Fujii}, M.~S., {Leigh}, N.~W.~C., and
  {Kumamoto}, J.: 2021b,
\newblock {\em \mnras} {\bf 504(1)}, 910

\end{thebibliography}

\end{document}